\documentclass[aps,twocolumn,aps,showpacs]{revtex4}
\usepackage{graphicx}
\usepackage{amsmath}
\usepackage{amssymb}
\usepackage{bm}

\newcommand{\be}{\begin{eqnarray}}
\newcommand{\ee}{\end{eqnarray}}
\newcommand{\ba}{\begin{array}}
\newcommand{\ea}{\end{array}}
\newcommand{\etal}{{\it et al.}}

\begin{document}

\title{Two impurity Kondo problem under Aharonov--Bohm and Aharonov--Casher Effects}

\author{Tomosuke Aono}
\affiliation{%
Department of Physics,
Ben-Gurion University of the Negev,
Beer-Sheva 84105, Israel}
\date{\today}
\pacs{%
72.15.Qm,   
73.23.-b,     
85.35.Ds,    
85.75.-d,     
}

\begin{abstract}
We investigate electron transport under the two impurity Kondo problem with
the Aharonov--Bohm and Aharonov--Casher effects.
These interference effects induce  the Ising-coupled Ruderman--Kittel--Kasuya--Yosida (RKKY) interaction. 
We  discuss the inter- and intra-site spin conductance as well as charge conductance in the Kondo and  the mixed-valence regimes using the slave boson mean field approximation.
\end{abstract} 

\maketitle

The Aharonov--Bohm (AB) effect~\cite{AB} plays a central role in 
interference effects in mesoscopic systems.
When the system is under the influence of the spin--orbit interaction (SOI),
an additional interference effect,  the Aharonov--Casher (AC) effect~\cite{AC,Meir,Mathur,Balatsky, Aronov,Qian} emerges;
An electron acquires a phase factor after passing through an AB ring because of
the interaction between the spin and  electric field through the ring ($z$-direction).
The electric field  can control electron transport through the ring~\cite{Nitta}. 
Several experiments~\cite{Morpurgo,Yau,Yang} discussed the interference effects under the SOI.
In a heterojunction, a structural inversion asymmetry of the confining potential near the junction induces
the Rashba SOI,
$\alpha (k_{y} \tau^{x} - k_{x} \tau^{y})$
with the coupling constant $\alpha$, which is controlled by the confinement electric field,
the wave vector $\vec{k}$ of electrons, and the Pauli matrix $\vec{\tau}$~\cite{Rashba}.
In a recent experiment~\cite{Koenig}, an interference pattern is clearly shown as
a function of electric and magnetic fields in an AB ring system with the Rashba SOI.
Many theoretical studies have been devoted to electron transport related to this situation~\cite{Souma,Frustaglia01,Meijer,Molnar,Capozza}. 

Since the AC effect induces spin-dependent phases, 
it can control spin states in certain geometries.
To demonstrate this,
we consider a coupled quantum dot system embedded in an AB ring under
the Ruderman--Kittel--Kasuya--Yosida (RKKY) interaction and Kondo effect. 
These two interactions compete with each other---
This is known as the two impurity
Kondo problem~\cite{Jones,Affleck,Jones89,Georges}.  
In a recent
experiment~\cite{Craig}, this competition was observed in a coupled quantum dot system.
Theoretical issues~\cite{Simon,Vavilov} related to this experiment were also discussed.
This competition is further investigated in gold grain quantum dots with
magnetic impurities in the leads~\cite{Heersche}.
The Kondo  effect under the AB effect was considered in
a coupled dot system~\cite{Izumida}, and  a triangle dot system~\cite{TKuzmenko}. 
In addition, Utsumi {\it et al.}~\cite{Utsumi} investigated
the AB flux dependent RKKY interaction and discussed
the two impurity problem in the perturbative regime.

In this paper, we investigate the two impurity Kondo problem under the AB and AC effects.
We will show these interference effects induce  the Ising-coupled RKKY interaction.
This model has been investigated in capacitive coupled quantum dot systems~\cite{Andrei, Garst}.
We also report spin conductance as well as charge conductance
in the Kondo and mixed-valence regimes using the slave boson mean field approximation~\cite{Coleman,Jones89}, demonstrating
how these effects control spin transport under the electron--electron interactions.

{\it RKKY interaction under SOI.---}
We first summarize the RKKY interaction under the SOI without the AC effect.
The spin-exchange interaction under the SOI  between two localized spins $\vec{S}_{i} \; (i=1,2)$ 
consists
the Heisenberg  interaction, $\vec{S}_1 \cdot \vec{S_2}$,
Ising (anisotropic) interactions such as $ S_{1}^{z} S_{2}^{z}$,
and the Dzyaloshinsky-Moriya (DM) interaction, $\vec{S_1} \times \vec{S_2}$.
The sum of these term can be  rewritten in a
compact form ~\cite{Shekhtman,Kavokin,Imamura}
\be
\label{eq:rkky_soonly}
H_{\rm ex} = J \vec{S}_1 \cdot \vec{S}_2(\theta),
\ee
where $J$ is the coupling constant, and $\vec{S}_2(\theta)$ denotes that
the spin quantization axis of $S_2$  is tilted from 
the axis of the first impurity with an angle $\theta$,
which  depends on the strength of the SOI. 
Equation (\ref{eq:rkky_soonly}) is derived as follows for the RKKY interaction,
$H_{\rm RKKY} = -\frac{J_{sd}^{2}}{\pi} {\rm Im} \int_{-\infty}^{E_{\rm F}} 
d\epsilon {\rm Tr} [ (\vec{S}_{1}\cdot \vec{\tau}_{1}) G(1,2,\epsilon+i0^{+})
(\vec{S}_{2}\cdot\vec{\tau}_{2})  G(2,1,\epsilon+i0^{+}) ]$
with the Green function $G(i,j,\epsilon)$ of conduction electrons under the SOI between the localized spin $i$ and $j$, and the s-d coupling constant $J_{sd}$.
$G$ is $2\times2$ matrix in the spin space.
We can eliminate the influence of the SOI  in $G$ to diagonalize it by
rotating the spin-matrix of conduction electrons~\cite{Shekhtman,Levitov}.
This procedure is equivalent to rotating the spin-matrix of the localized spin instead of one of the conduction
electrons.  Then interaction Hamiltonian can be written
\be\label{eq:rkky_sd}
H_{\rm RKKY}=
\sum_{\sigma,\sigma'=\pm \atop a,b=x,y,z}
 \tau_{\sigma \sigma'}^{a} \tau_{\sigma' \sigma}^{b} S_{1}^{a} S_{2}^{b}(\theta)
f(1,2),
\ee
where $\theta = 2 m \alpha R/ \hbar^2$, with electron mass $m$,
distance $R$ between the two impurities.
$f(1,2)$ is the RKKY function determined by $R$, $J_{\rm sd}$, and
the effective Fermi wavelength $q_F =\sqrt{2 m E_F/\hbar^2 + (m \alpha/\hbar^2)^2}$~\cite{Imamura}.
After taking the sums, we obtain $H_{\rm RKKY}$ in the form of Eq.~(\ref{eq:rkky_soonly}).
The amplitude and sign of $J$ depends on $q_{F} R$~\cite{Imamura}.

We attach external leads to the impurities to induce the Kondo effect,
$
H_{\rm Kondo} = \sum_{k,\sigma \atop i=1,2} \epsilon_{k} \bar{c}_{k i \sigma}
c_{k i \sigma} 
 + \sum_{i=1,2} J_{i} \vec{S}_i \cdot \vec{\tau},
$
where $c_{k i \sigma}$ is the annihilation operator  of conduction electrons
with energy $\epsilon_{k}$, and $J_{i}$ is the s-d coupling constant.
We can rotate the quantization axis of conduction electrons in the lead $2$  with $\theta$ because the kinetic term of the conduction electrons in the leads
is invariant under rotation.
Therefore,
the two-impurity model, $H_{\rm Kondo}+H_{\rm RKKY}$, is the same as the one without the SOI,
except that $J$ is modified.
Note that in general, $H_{\rm Kondo}$ includes the Kondo effect from the ring with  $J_{\rm sd}$.

{\it RKKY interaction under AB and AC effects.---}
\begin{figure}
\includegraphics[width=8cm]{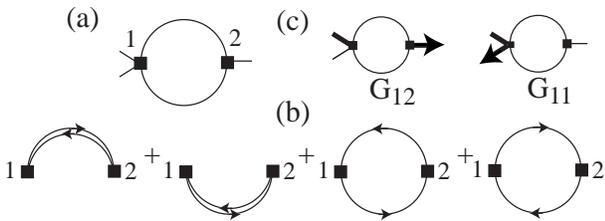}
\caption{%
(a) Two impurity device geometry. Two external leads are attached to the site 1.
(b) Four possible rounds of electrons contributing to the RKKY interaction.
(c) Inter- and intra-site electron transport.
\label{fig:geometry}}
\end{figure}
The AB and AC effects make a qualitative difference in
the above situation.
We consider the two impurity system in a ring with a radius $r$, connected to external leads
as depicted in Fig.~\ref{fig:geometry}(a).
The magnetic and electric fields pierce the ring.
When an electron passes through the upper/lower branch ($n=\pm1$),
with spin $\sigma$, traveling from the impurity $1$ to $2$,
the electron acquires the phase factor of $\exp [ i n  \Phi_{\rm t}(\sigma)]$
with $\Phi_{\rm t}(\sigma) =\Phi + \sigma \Phi_{\rm c}$~\cite{Balatsky},
where $\Phi$ is the AB phase.
The factor of $\Phi_c$ comes from the AC effect,
$
\Phi_{\rm c} =   -\pi/2(1-\cos \theta ) - \pi r \frac{m \alpha}{\hbar^2} \sin \theta 
$
with $\tan \theta = - 2\alpha m r/\hbar^2$~\cite{Qian}.
In the following, we treat $\Phi$ and $\Phi_c$ as external parameters.
 
To calculate the RKKY interaction,
we need to take into account the four possible rounds of  electrons as depicted in Fig.~\ref{fig:geometry}(b).
For the AB effect, this procedure has been discussed in Ref.~\cite{Utsumi}.
Note that the statistics of the RKKY coupling constant in disordered conductors has been discussed in Refs.~\cite{Vavilov} and \cite{Zyuzin}.
The result is, instead of Eq.~(\ref{eq:rkky_sd}),
$
 H_{\rm RKKY}=
\sum_{\sigma,\sigma'=\pm \atop a,b=x,y,z}
 \cos\Phi_t(\sigma) \cos \Phi_t(\sigma') 
\tau_{\sigma \sigma'}^{a} \tau_{\sigma' \sigma}^{b} S_{1}^{a} S_{2}^{b}
f(1,2).
$
We trace out the spin operators of the conduction electrons to obtain
$
H_{\rm RKKY}
=
J \cos \left( \Phi - \Phi_{c} \right) \cos \left( \Phi + \Phi_{c} \right)
(S_{1}^{x} S_{2}^{x} + S_{1}^{y} S_{2}^{y})  
+
J /2 \left[
 \cos^{2} \left( \Phi- \Phi_{c} \right) 
 +
\cos^{2} \left( \Phi+ \Phi_{c} \right) 
\right]
S_{1}^{z} S_{2}^{z}
$.
We have  disregarded the $\theta$ dependence in $\vec{S}_{2}$.
Although the coupling constant of $S_1^{x} S_2^{x} \;(S_1^{y} S_2^{y})$ can change sign for
certain values of $\Phi$ and $\Phi_c$,
this can be absorbed by a transformation $S_2^{x/y} \rightarrow -S_2^{x/y}$.

The RKKY interaction is eventually written in the form
\be\label{eq:rkky-phase}
H_{\rm RKKY} = J_1 \vec{S}_1 \cdot \vec{S}_2 + J_2 S_{1}^{z} S_{2}^{z}
\ee
with $J_1 = J |\cos \left( \Phi - \Phi_{c} \right) \cos \left( \Phi + \Phi_{c} \right) |$
and
$J_2 = J /2 [
 \cos^{2} ( \Phi - \Phi_{c} )  + \cos^{2} ( \Phi + \Phi_{c} ) ]- J_1$.
The Ising exchange term appears in addition to
the conventional Heisenberg term.
The coupling constants are tuned by the external magnetic and electric fields
via the AB and AC effects.
Note that this result originates from the interference effects.
Thus,
if one of the branches is disconnected, $J_2$ returns to zero even
when the SOI is nonzero.
When the phase factors for the upper and lower arms are different,
the $z$-component of the DM type interaction,
 $(\vec{S}_{1} \times \vec{S}_{2})^{z}$, can be induced.
This term, together with the Heisenberg term, is then expressed by
$\exp (i \eta) S_1^{+} S_2^{-} + {\rm h.c.}$, with  spin raising and lowering operators
$S^{\pm}$ and a
certain phase factor
$\eta$. 
This $\eta$ can be removed by the transformation
$S_2^{\pm} \rightarrow \exp ( \pm i\eta) S_2^{\pm}$. 
Thus the $z$-component of the DM  interaction is irrelevant for the problem considered here.

{\it Model and approximations.---}
To discuss the transport,
we start from the two impurity Anderson model~\cite{Coleman,Jones89,Georges,Aono}
with the RKKY interaction:
$H = 
\sum_{k, i, \sigma} \epsilon_{k} \bar{c}_{k i \sigma} c_{k i \sigma}  
+ V_g  \sum_{i, \sigma}  \bar{c}_{i \sigma} c_{i \sigma}
+ V_c  \sum_{\sigma} 
\left(  \bar{c}_{1 \sigma} c_{2 \sigma} + {\rm h.c.} \right)
+ H_{\rm RKKY} +
U \sum_{i} n_{i \uparrow} n_{i \downarrow} +
V \sum_{k,i,\sigma} \left( \bar{c}_{k i \sigma} c_{i \sigma} + {\rm h.c.}  \right),
$
where
$c_{i\sigma}$ is the annihilation operator of  site $i$ electrons 
with spin $\sigma$, 
$n_{i\sigma} = \bar{c}_{i\sigma} c_{i\sigma}$, and
$V_{g}$ is the gate voltage~\cite{vgcomment}.
Note that the spin quantization axis is different between sites 1 and 2
as in the RKKY interaction, and $\sigma=\pm$ spin state is defined by the local axis.
We have introduced
the direct tunneling coupling $V_{c}$.
The spin state follows the local quantization axis and is unchanged during the tunneling between the sites.
We assume  $J > 0$ in $H_{\rm RKKY}$.
We also assume the on-site Coulomb energy $U \rightarrow \infty$,
which allows to use the slave boson representation:
$c_{i\sigma} = \bar{b} f_{i\sigma}$ with the slave boson operator $b$ and pseudo
fermion operator $f_{i\sigma}$
with  the constraint term,
 $H' = \lambda \sum_{i} ( \sum_{\sigma} \bar{f}_{i\sigma} f_{i\sigma} + \bar{b} b - 1)$.
We adapt the mean field theory~\cite{Coleman,Jones89,Aono},
introducing an extra mean field $m = J_{2} \langle S_1^z \rangle = - J_{2}\langle S_2^z \rangle$ for the Ising anti-parallel interaction. 
We discuss the choice of $m$ later.
We have disregarded the Kondo effect from the ring since it does not change the conclusions below.
We also discuss this point  later.

The model is now reduced to the two site non-interacting model under the AB and AC effects with 
the effective energy scales as follows: $H = 2\lambda (b^{2}-1) +  \kappa^{2}/J_{1} +  m^{2}/J_{2}+ H_{0}$, where
\be
H_{0} = \sum_{\sigma}
[\bar{f}_{1 \sigma}, \bar{f}_{2 \sigma}]
\hspace{-1mm}
\left[
\begin{array}{cc}
\tilde{E} + m +  i \tilde{\Delta} & V_c (\sigma) \\
V_c (\sigma) &  \tilde{E} - m +  i \tilde{\Delta} 
\end{array}
\right]
\hspace{-1.5mm}
\left[
\begin{array}{cc}
f_{1 \sigma}  \\
f_{2 \sigma}
\end{array}
\right]
\ee
with
the effective energy level $\widetilde{E}=V_{g}+\lambda$, and
the effective site-lead coupling $\tilde{\Delta}=b^2 \Delta$~\cite{Coleman,Jones89},
where
$\Delta = \pi \rho  |V|^2$ with the density of states $\rho$ of the lead electrons at
the Fermi energy.
Since there are two possible branches to reach from one site to the other site with different phase factors due to the AB and AC effects,
the effective coupling constant
$V_c (\sigma)$ between the sites depends on the phases: 
\be
 V_c (\sigma) =  \left( \kappa + V_c \frac{\tilde{\Delta}}{\Delta} \right) \cos \Phi_t(\sigma),
\ee
where $\kappa$ is the spin-singlet mean field parameter
due to the Heisenberg  term~\cite{Coleman,Jones89}.
The cosine factor represents the interference of the hopping term between the upper and lower branches and it induces the spin dependence.
Note that if one of the branches is disconnected, the spin dependence disappears. 
We have disregarded multiple backscattering inside the ring~\cite{opencondition}.
We solve the self-consistent equations for $\widetilde{E}$, $\widetilde{\Delta}$, $\kappa$,
and $m$ for given values of $V_g$, $\Phi$, and $\Phi_{c}$,
choosing the lowest energy solution among the possible solutions.
We calculate the inter- and intra- site conductance, as depicted in Fig.~\ref{fig:geometry}(c):
 $G_{12 \sigma} = 4 e^{2}\Delta^{2}/h  |\langle c_{1 \sigma} \bar{c}_{2 \sigma} (\omega=0)\rangle|^{2}$,
 $G_{11 \sigma}=  -2 e^{2} \Delta /h {\rm Im} \langle c_{1 \sigma} \bar{c}_{1 \sigma}(\omega=0)\rangle $
 with the retarded Green function $\langle c_{i \sigma} \bar{c}_{j \sigma}(\omega) \rangle$ between
 the site $i$ and $j$.
Note that $G_{12}$ has a single spin index because of $V_{c}(\sigma)$.
 To measure $G_{11\sigma}$, we need an extra lead for the site 1
 as depicted in Fig.~\ref{fig:geometry},
assuming the equal dot-lead couplings.

{\it Phase-controlled spin state---}
\begin{figure}
\includegraphics[width=5cm]{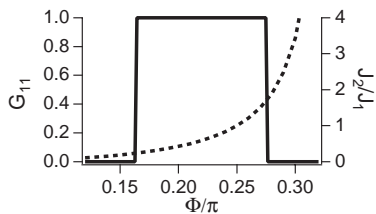}
\caption{%
 $G_{11c}$ (solid line) and $J_{2}/J_{1}$ (dotted line)  as a function of $\Phi$
at $\Phi_{c}=\pi/6$, $V_g/\Delta = -1.5$,  and  $V_c= 0$,
In Figs.~\ref{fig:tik_ac}-\ref{fig:mixed-valence}, $J/\Delta=2.5\times10^{-2}$, and the unit of $G$ is $2e^{2}/h$.
\label{fig:tik_ac}}
\end{figure}
First, we demonstrate the spin state in the ring can be tuned by
the AB and AC effects.
In  Fig.~\ref{fig:tik_ac},
$G_{11}= G_{11+}= G_{11-}$ and $J_{2}/J_{1}$ are plotted as a function of $\Phi$ for $\Phi_{c}=\pi/6$ and $V_{c}=0$.
When $\Phi$ is small, the spin-singlet state due to the Heisenberg exchange is the ground state.
As $\Phi$ increases, the state changes to the Kondo state,
resulting in  the finite conductance.
As $\Phi$ increases further, when $J_{2} > J_{1}$, it becomes the Ising state with zero conductance.
When $\Phi \pm \Phi_{c} = \pi/2 n$ ($n$ is an integer),
$J_{1}$ is zero while $J_{2}$ is finite; the Ising-coupled two impurity model is realized.
The model has been investigated
in capacitive coupled quantum dot systems and
this model exhibits a quantum phase transition~\cite{Andrei,Garst}.
The system presented here is another realization of this model using
the phase-coherent phenomena.

{\it Spin transport under two-impurity model.---}
\begin{figure}
\includegraphics[width=8cm]{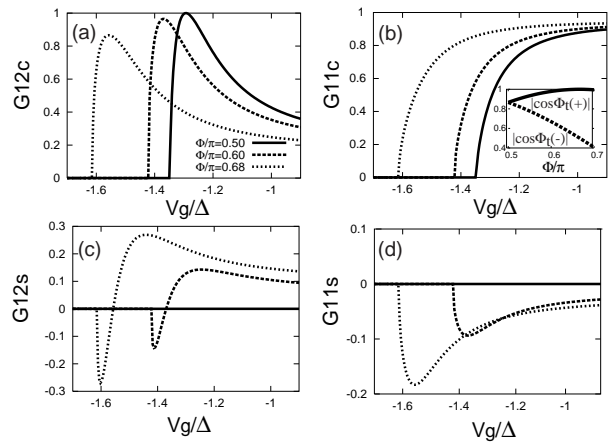}
\caption{%
(a) $G_{12c}$, (b) $G_{11c}$, (c) $G_{12s}$, and (d) $G_{11s}$  vs. $V_g$ for $\Phi/\pi =
0.5,0.6,0.68$ and $\Phi_c=\pi/3$.
Inset in (b): $|\cos(\Phi \pm \Phi_c)|$  [solid (broken) line] vs. $\Phi/\pi$ .
We add $V_c/\Delta= 0.3$ to have smooth sharp peaks. See the discussion later.
\label{fig:tik_spin}}
\end{figure}
Next, we discuss the charge/spin conductance,
$G_{12c/s} \equiv G_{12+} \pm G_{12-}$ and $G_{11c/s} \equiv G_{11+} \pm G_{11-}$.
In Fig.~\ref{fig:tik_spin},
$G_{12c/s}$ and $G_{11c/s}$ are plotted as a function of  $V_g$  for several values of $\Phi$ for
$\Phi_c=\pi/3$.
In this figure, 
$J_{2}<J_{1}$ and $m=0$;
the Heisenberg coupling dominates the RKKY interaction.
The curves of $G_{12c}$ show single peak structures, while
the curves of $G_{11c}$ show single step structures.
The curves of $G_{12s}$ show double extremum structures, with
one maximum and one minimum.
The curves of $G_{11s}$ on the other hand,  show  single peak structures.
This means that the inter- and intra-sites spin current can flow in opposite directions
for a certain range of $V_g$.

When $V_g$ is high, the system is dominated by the Kondo effect while 
the Heisenberg exchange is less prominent.
As $V_{g}$ decreases, the Kondo singlet and spin-singlet states coexist,
where $V_{c}(\sigma)$ starts to develop.
This results in the peak of $G_{12c}$~\cite{Georges,Aono}.
On the other hand,
the same effect suppresses the intra-site conductance,
resulting in the step of $G_{11c}$.
In the inset of Fig.~\ref{fig:tik_spin} (b),
$|\cos(\Phi + \sigma \Phi_c)|$ is plotted as a function of
$\Phi$.
The spin transport is obtained when $\Phi/\pi > 0.5$,
where the $V_{c}(+) > V_{c}(-)$.
When $V_{g}$ is high, $G_{12s}$ is determined by $V_{c}(+)$, and $G_{12s}>0$ while $G_{11s}$ is determined by $V_{c}(-)$, and  $G_{11s} < 0$.
When $V_{g}$  decreases, $V_{c}(+)$ becomes larger so that the up spin level is away from the Fermi level. Then $V_c(-)$ term is the main contribution in
$G_{12s}$,
resulting in  $G_{12s} < 0$.

\begin{figure}
\includegraphics[width=8cm]{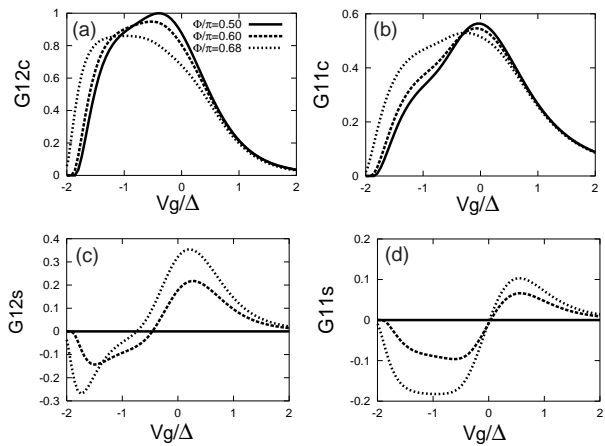}
\caption{%
(a)$G_{12c}$, (b) $G_{11c}$, (c) $G_{12s}$, and (d)  $G_{11s}$ vs. $V_g$ for $\Phi/\pi =
0.5,0.6, and 0.68$ for $\Phi_c=\pi/3$ and $V_c/\Delta=1.5$.
\label{fig:mixed-valence} }
\end{figure}

{\it Mixed-valence regime.---}
Next, we consider 
the mixed-valence regime.
We focus on the case of $V_{c} / \Delta = 1.5$ to clarify the role of the spin correlations in the previous results.
Figures~\ref{fig:mixed-valence}(a)-(d) show
$G_{12c/s}$ and $G_{11c/s}$.
The curves of $G_{12c/s}$ are qualitatively similar to those in Fig.~\ref{fig:tik_spin}.
On the other hand, $G_{11c/s}$ are qualitatively different;
$G_{11c}$ show  a peak instead of  a step structure, and 
$G_{11s}$ is qualitatively similar to $G_{12s}$
unlike the one in the Kondo regime.
When $V_{c}/\Delta > 1$, $V_{c}(\sigma)$ is determined by
$V_{c} \widetilde{\Delta}/\Delta$.
This means the peak structures come
the splitting of the bonding and anti-bonding states and
the occupation in the sites rather than
the competition between the Kondo and RKKY correlations~\cite{Aono}.
The peaks in $G_{12/11c}$  appear when
the bonding (lower) level  crosses the Fermi level in the leads.
Since $V_{c}(+) > V_{c}(-)$, 
the up spin state first reaches at the Fermi level,
resulting in $G_{12/11s} > 0$.
When $V_{g} < 0$, the up spin levels are away from the Fermi level,
the down spin levels dominates the spin transport, resulting in $G_{12/11s}<0$.

We should discuss two effects in the Kondo regime, when $V_{c}$ is finite;
the Kondo effect from the ring and the fluctuations from the mean field approximation~\cite{Coleman}.
The Kondo effect from the ring will induce an extra site-lead tunneling coupling 
$\Delta' \sim \rho_{\rm r} V_{c}^{2}$ with
the density of states $\rho_{\rm r}$ of the ring.
Then the Kondo temperature $T_{K}\propto \exp(\pi V_{g}/(\Delta+\Delta'))$.
When $V_{g}$ is normalized by $\Delta + \Delta'$, the result is the same because
the competition between $T_K$ and $J$ is the central part of the two-impurity problem.
The fluctuations around the mean field induce additional RKKY interactions~\cite{Coleman}; $V_{c}^{2}/V_{g}$ coupling is induced.
When the RKKY interaction is dominated by the Heisenberg exchange,
it eventually  modifies $V_{c}(\sigma)$, which will explain the results as those
in Fig.~\ref{fig:tik_spin}.
When the Ising coupling dominates the interaction, 
this coupling lifts the degeneracy of the Ising doublet state;
the fluctuations of $m$ are large and the mean field approximation becomes invalid.
More quantitative analysis is required in the regime.

In conclusion,
we investigated the two impurity Kondo problem
under the AB and AC effect.
The AC effect induces the Ising-coupled RKKY coupling.
These interference effects can control the spin states
as well as spin transport,
which is qualitatively different between the Kondo and mixed-valence regimes.

We would like to thank 
A.~Aharony, Y.~Avishai, O.~Entin-Wohlman,  Y.~Meir, R.~Shaisultanov, K.~Takahashi, and R.~Tasgal
for comments and discussions.



\begin{thebibliography}{99}
\bibitem{opencondition}
This means the ring is widely opened to the leads.
\bibitem{vgcomment}
$V_{g}$ is defined relative to the electric field in the ring to
keep the confinement of the sites.
\end{thebibliography}
\end{document}